\documentclass[letterpaper]{article} 
\usepackage{aaai25}  
\usepackage{times}  
\usepackage{helvet}  
\usepackage{courier}  
\usepackage[hyphens]{url}  
\usepackage{graphicx} 
\usepackage{natbib}  
\usepackage{caption} 
\usepackage{algorithm}
\usepackage{algorithmic}

\usepackage{newfloat}
\usepackage{listings}
\DeclareCaptionStyle{ruled}{labelfont=normalfont,labelsep=colon,strut=off} 
\floatstyle{ruled}
\newfloat{listing}{tb}{lst}{}
\floatname{listing}{Listing}
\title{``Accessibility People, You Go Work on That Thing of Yours over There'': Addressing Disability Inclusion in AI Product Organizations}
\author{
    Sanika Moharana\textsuperscript{\rm 1}, 
    Cynthia L. Bennett\textsuperscript{\rm 2}, 
    Erin Buehler\textsuperscript{\rm 3}, 
    Michael Madaio\textsuperscript{\rm 2}, 
    Vinita Tibdewal\textsuperscript{\rm 3}, 
    Shaun K. Kane\textsuperscript{\rm 2}}
\affiliations{
    \textsuperscript{\rm 1} Carnegie Mellon University\\
     \textsuperscript{\rm 2}Google Research\\
      \textsuperscript{\rm 3}Google Inc.\\

    smoharan@andrew.cmu.edu, clbennett@google.com, ebuehler@google.com, madaiom@google.com, vtibdewal@google.com, shaunkane@google.com
}

\usepackage{bibentry}

\begin{document}

\maketitle

\begin{abstract}
The rapid emergence of generative AI has changed the way that technology is designed, constructed, maintained, and evaluated. Decisions made when creating AI-powered systems may impact some users disproportionately, such as people with disabilities. In this paper, we report on an interview study with 25 AI practitioners across multiple roles (engineering, research, UX, and responsible AI) about how their work processes and artifacts may impact end users with disabilities. We found that practitioners experienced friction when triaging problems at the intersection of responsible AI and accessibility practices, navigated contradictions between accessibility and responsible AI guidelines, identified gaps in data about users with disabilities, and gathered support for addressing the needs of disabled stakeholders by leveraging informal volunteer and community groups within their company.  Based on these findings, we offer suggestions for new resources and process changes to better support people with disabilities as end users of AI.

\end{abstract}

\renewenvironment{quote}
  {\small\itshape\list{}{\rightmargin\leftmargin}\item\relax}
  {\endlist}

%

\section{Introduction}

The tech industry is increasingly orienting itself around developing and propagating generative AI (genAI) models, integrating them into mainstream applications and operating systems. Simultaneously, software companies and community organizations are raising concerns about AI’s potential impacts on society, including effects on education \cite{AnthropicEducationReport, openaistaffCollegeStudentsChatGPT2025, holmes2023ethics} and employment \cite{demirciResearchHowGen2024,kochharWhichUSWorkers2023}. Specific attention has also been placed on AI’s negative impacts on marginalized groups. These concerns are marked by high profile examples of AI-based discrimination through race, ethnicity, gender, and other sociodemographics, such as the racist labeling of photos \cite{hernGooglesSolutionAccidental2018,macFacebookApologizesAI2021}. In response, researchers have conducted various studies to understand and mitigate these harms \cite{buolamwiniGenderShadesIntersectional2018}, while companies have implemented responsible AI (RAI) guidelines (e.g., \citealt{microsoftResponsibleAIPrinciples,googleAIPrinciples}) and have even withdrawn products due to problematic AI output \cite{grantGoogleSaysIt2024}.

Journalists covering AI impacts have also reported on AI’s role in perpetuating discrimination against people with disabilities (e.g., \citealt{smithAutomatingAbleism2024,lecherHealthcareAlgorithmStarted2018}). Advances in technology often lead to new barriers for people with disabilities (hereafter PWD) \cite{williamsonAccessibleAmericaHistory2019,holmesMismatchHowInclusion2020}. Researchers have identified a breadth of negative impacts downstream from AI applications, in domains from housing to employment to education \cite{trewinConsiderationsAIFairness2019,aboulafiaBuildingDisabilityInclusiveAI2025}. For example, the user interface for AI chatbots may be inaccessible \cite{torresAccessibilityChatbotsState2019},  and AI resume reviewers have been shown to demonstrate ableist bias  \cite{glazkoIdentifyingImprovingDisability2024}. These concerns are effectively expressed by \citet{trewinConsiderationsAIFairness2019} stating, “the prospects of AI for PWD are promising yet fraught with challenges.”

Simultaneously, widespread reputable digital accessibility best practices such as the Web Content Accessibility Guidelines (WCAG) do not yet cover novel interactions and outputs generated by emergent AI systems \cite{montgomeryW3CAccessibilityGuidelines2024}. For example, they do not provide guidance in ensuring accessibility of model outputs, which are often the defining novel features of genAI tools like large language models (LLMs), nor do they advise developers on mitigating disability representation biases pervasive in these outputs \cite{gadirajuWouldntSayOffensive2023}. Thus, there is an opportunity to understand how AI practitioners currently approach disability inclusion in their work, what gaps remain in AI product organizations, and how we should ensure future AI systems are inclusive of users with disabilities.
Our research is guided by the following questions:

\begin{itemize}
\item RQ1: How are AI practitioners encountering disability and accessibility exclusion in genAI product development?
\item RQ2: What obstacles do AI practitioners encounter in identifying and handling disability and accessibility?
\item RQ3: What strategies do AI practitioners leverage to successfully handle disability and accessibility exclusion?
\end{itemize}

To investigate these questions, we conducted a semi-structured interview study with 25 AI practitioners at a large software company, to discuss how AI is changing the work that they do, and particularly how their work may or may not be inclusive of disabled people. Participants held a range of roles, including engineers, researchers, UX practitioners, and people working in the responsible AI space (e.g., safety analysts) (see Table \ref{table:participants}). We found:

\begin{itemize}
\item Decisions that may affect users with disabilities may appear across a variety of work contexts that lie outside the bounds of traditional accessibility roles, including model development and safety evaluation;
\item Existing practices around AI safety and responsibility evaluation may emphasize specific identity categories, and conversely pay less attention to users with disabilities as stakeholders;
\item Existing methods for evaluating and verifying accessibility may be ineffective, due to not being used, not being prioritized, or lacking in guidelines relevant to AI.
\end{itemize}

These findings point to a broader need to reconsider and understand accessibility and disability inclusion in the context of AI. Specifically, this work contributes:

\begin{itemize}
\item The first study of how AI practitioners navigate disability and accessibility in generative AI product development;
\item Identification of gaps in how problems impacting users with disabilities may be overlooked in AI;
\item Suggestions from practitioners for improving how users with disabilities are accounted for in the AI development and release process.
\end{itemize}

\section{Related Work}

\subsection{Studies of Accessibility Practitioners}

While the International Association of Accessibility Professionals has issued over 8500 certifications~\cite{iaapIAAPHome}—there exists relatively limited scholarship on the work they perform. Early studies of web professionals \cite{lazarImprovingWebAccessibility2004,rossonDesigningWebRevisited2005}, product designers \cite{vanderheidenBarriersIncentivesFacilitators1998}, and UX experts \cite{putnamHowProfessionalsWho2012,yesiladaExploringPerceptionsWeb2015} noted that adherence to accessibility standards was low, citing barriers including confusing guidelines, lack of time, training, and support from managers.

Other studies have examined professional practices around accessibility within specific communities or regions (e.g., \citealt{freireSurveyAccessibilityAwareness2008, almerajUnderstandingMindsetsSkills2023, inalWebAccessibilityTurkey2019, guptaWebAccessibilityWeb2021}). Across these examples, researchers identified challenges including insufficient time, educational resources, and knowledge of available accessibility tools. Similar challenges persist today, although the rise of social networking communities has provided additional peer support and learning opportunities \cite{huqA11yDevUnderstandingContemporary2023}.

More recent studies have focused on the motivations and work conducted by accessibility professionals. Shinohara et al. \citeyear{shinoharaWhoTeachesAccessibility2018}  found that about half of the US education institutions they reached out to had at least one accessibility educator, but their work was sometimes limited by unclear objectives and insufficient curriculum development support. Leitner et al. \citeyear{leitnerWebAccessibilityImplementation2016} surfaced three motivators of industry accessibility professionals: economic (e.g., increasing the customer base), demonstrating social commitments, and technical (e.g., improving site design). However, Azenkot et al. \citeyear{azenkotHowAccessibilityPractitioners2021} and Pereira and Duarte \citeyear{seixaspereiraEvaluatingMonitoringDigital2025} surfaced the advocacy required of accessibility professionals; \cite{azenkotHowAccessibilityPractitioners2021}  noted that educating others was a core job function. Challenges cited by \cite{seixaspereiraEvaluatingMonitoringDigital2025}  included a gap between user needs and accessibility compliance and tooling needs, especially to test mobile web accessibility.

Our study extends prior work by exploring how accessibility and disability are considered during the development and evaluation of emerging AI technologies, by centering perspectives from professionals at the frontlines–those who work on accessibility and AI.

\subsection{Generative AI and Disability}
Research exploring the impacts of generative AI on PWD has grown significantly in recent years \cite{alshaigyForgottenAgainAddressing2024}. One strand of research has identified AI harms experienced by PWD \cite{whittaker2019disability,kirmerDisabilityAccessibilityAI2024,newman-griffisDefinitionDrivesDesign2023}. Reports classify harms in a variety of domains including employment, education, safety, and healthcare \cite{trewinConsiderationsAIFairness2019,ainowinstituteDisabilityBiasAI2019,aboulafiaBuildingDisabilityInclusiveAI2025}. Hutchinson et al. \citeyear{hutchinsonSocialBiasesNLP2020} and Venkit et al. \citeyear{venkitStudyImplicitBias2022} surfaced negative terms and sentiments language models associated with disability labels. These misrepresentations extend to generated stories and images, which perpetuated longstanding disability tropes (e.g., disabled people as inspirational or cure-seeking) \cite{gadirajuWouldntSayOffensive2023,mageeIntersectionalBiasCausal2021,venkitAutomatedAbleismExploration2023, bianchiEasilyAccessibleTexttoImage2023,mackTheyOnlyCare2024}. Disability harms may also reverberate through algorithmic decision-making, such as in a resume ranking task, ChatGPT rated more qualified applicants with disability language lower than resumes with no disability language \cite{glazkoIdentifyingImprovingDisability2024}. Further, toxicity classifiers misjudged ableism as compared to disabled people, demonstrating that content filters may not address disability-related online safety \cite{phutaneColdCalculatedCondescending2025}.

However, rapid improvements in genAI have allowed researchers to make progress on a variety of accessibility challenges , including image description \cite{penuelaUnderstandingUseMLLMEnabled2025,xieEmergingPracticesLarge2024}, image generation \cite{huhGenAssistMakingImage2023}, supporting writing for people with dyslexia \cite{goodmanLaMPostDesignEvaluation2022}, enhancing communication \cite{valenciaLessTypeBetter2023,weinbergWhySeriousExploring2025,caiUsingLargeLanguage2023,jangItsOnlyThing2024}, and supporting better captions for Deaf and hard-of-hearing users \cite{wuCARTGPTImprovingCART2024}. These tools offer real-time, bespoke outputs, enabling customization idealized for users whose needs are often unmet by mainstream designs. For example, generative image descriptions allow blind and low vision users to get visual information on-demand and tailored to their information needs, via visual question answering (VQA) interfaces \cite{bemyeyesIntroducingBeMy2023}. Generative communication supports are responsive to open ended inputs, allowing users to engage naturally \cite{goodmanLaMPostDesignEvaluation2022, valenciaLessTypeBetter2023, wuCARTGPTImprovingCART2024}. These promising developments situate PWD as key beneficiaries of genAI applied to longstanding accessibility challenges.

We interviewed AI practitioners working across a wide array of projects–from ensuring PWD can benefit from genAI integrated into existing, mainstream products, as well as developing accessibility-specific products. We report back on concerns and challenges they experienced related to AI and disability in these pursuits.

\subsection{Studies of Responsible AI Practice}
Researchers are not only concerned about disability-based bias, but about AI’s potential adverse impacts on individuals and society at large. Much of this work has coalesced around developing guidance for and evaluating fairness, accountability, and transparency, often condensed under the umbrella of "Responsible AI " (RAI).  In industry, commitments to RAI are often documented via a set of public principles (e.g., \citealt{microsoftResponsibleAIPrinciples,googleAIPrinciples}). RAI as an emerging discipline has been shaped by formative work such as \textit{Gender Shades} \cite{buolamwiniGenderShadesIntersectional2018}, which identified intersectional disparities in commercial gender classification systems along gender and skin tone dimensions, and by documenting and classifying sociotechnical harms \cite{shelbySociotechnicalHarmsAlgorithmic2023}.

Recent work has explored the emerging job roles around RAI \cite{rismani2023does}. This includes documenting the work that professionals do to ensure that AI systems are fair \cite{madaioAssessingFairnessAI2022,dengInvestigatingPracticesOpportunities2023,bermanScopingStudyEvaluation2024}, how UX practices are being adapted into RAI practices \cite{wangDesigningResponsibleAI2023}, and how professionals learn about RAI "on the job" \cite{madaioLearningResponsibleAI2024}. 

RAI research has primarily focused on a subset of identity categories such as gender, race, and ethnicity \cite{tahaeiSystematicLiteratureReview2023}, but the aforementioned examples of disability bias have not systematically impacted the field. In this study, we discuss how disability-related concerns appear in the work of AI professionals.

\section{Method}
We conducted 60-minute interviews with 25 AI practitioners at a large software company. They covered participants’ current work practices, how their work intersects with AI, and how their work intersects with PWD. This protocol was approved via our organization’s research review process. Participants completed an informed consent process for participating in the study and were provided with incentives of the equivalent of a \$60 USD gift card.

\subsection{Research Site}
We interviewed participants at a large, multinational technology company which produces hardware products, software products, and services. In recent years, the company has increased its focus on developing AI models and integrating AI-based features into its products. Some participants served on vertical teams (e.g., for a specific product or initiative) while others served on horizontal teams (i.e., providing a central resource or oversight across a range of products). Regardless of the specific team, participants all had access to some company-wide resources including trainings; their everyday work was typically guided by their own team's policies and practices, while also engaging with company-wide policies and processes, especially around product launches and compliance reviews.
\subsection{Researcher Positionality}
Our team comprised six researchers at the technology company from which participants were recruited. All were trained in human-computer interaction, some have experience with RAI, and a majority were also experienced accessibility researchers.

\subsection{Recruitment}
Since we were interested in learning AI practitioners’ experiences with disability inclusion and accessibility holistically irrespective of job title, we recruited participants holding a variety of roles, including researchers, data scientists, product managers, user experience practitioners, and AI safety analysts.  Recruitment announcements were emailed to both general-purpose and accessibility-focused discussion groups. All eligible participants had previously worked on genAI product development; we also asked about their work experiences related to accessibility, though such experience was not required. We conducted snowball sampling following suggestions made by other participants.

We received 81 completed screeners. From these, we selected 25 participants based on several criteria. First, we intended to recruit participants both from more traditional UX and accessibility roles (e.g., UX researchers, which is a common job role for accessibility practitioners in this company), as well as participants who did not have a UX or accessibility role but who may have encountered issues that could impact users with disabilities (e.g., data scientists and AI safety analysts). Second, we recruited participants who worked in different parts of the company or on different products. Finally, we tried to recruit a sample with diverse ethnic and gender identities.

\subsection{Participants}
Twenty-five participants took part in the study. Their professional roles are described in Table \ref{table:participants}. All participants worked in the US, except for P10, who worked for an international team (led by a manager in the US) but who lived in India. Further demographics are listed below
\footnote{Ethnicity of Participants: 7 Asian, 1 Black or African American; 1 Indian; 2 Middle Eastern or North African; 1 Mixed Race; 11 White; 1 White, Hispanic, Latino, or Spanish origin; 1 White, Middle Eastern or North African. Gender: 10 Men; 13 Women; 2 Non-binary. Disability Identity: 7 Disabled; 16 Non-Disabled; 2 Prefer not to say.}.

\begin{table*}[t]
\centering
\small
\begin{tabular}{l|lcccc}
ID  & Role  & \multicolumn{1}{l}{Years in role} & \multicolumn{1}{l}{AI experience?} & \multicolumn{1}{l}{Responsible AI experience?} & \multicolumn{1}{l}{a11y experience?} \\
\hline
P1  & Data Scientist & 1-2 years  & yes & yes  & no \\
P2  & Data Scientist & 5-9 years  & yes & no & no \\
P3  & Data Scientist & 3-4 years  & yes & yes  & no \\
P4  & Data Scientist & 5-9 years  & yes & no & no \\
P5  & Data Scientist & 1-2 years  & yes & yes  & no \\
P6  & Safety Analyst & {\textless{}1 year}  & yes & yes  & yes \\
P7  & Safety Analyst & {\textless{}1 year} & yes & yes  & no \\
P8  & Sales Analyst  & 1-2 years  & yes & yes  & no \\
P9  & UX Researcher  & 10+ years  & yes & unsure & no \\
P10 & UX Manager & 3-4 years  & yes & no & yes \\
P11 & Safety Analyst & 3-4 years  & yes & yes  & no \\
P12 & UX Manager & 1-2 years  & yes & yes  & yes \\
P13 & Safety Analyst & \textless{}1 year  & yes & yes  & yes \\
P14 & {Engineering Manager} & Not specified & yes & yes  & yes \\
P15 & {UX Researcher} & 1-2 years  & yes & no & yes \\
P16 & {Data Scientist}  & 3-4 years  & yes & yes  & no \\
P17 & {Product Manager} & 1-2 years  & yes & yes  & yes \\
P18 & {Researcher}  & 3-4 years  & yes & yes  & yes \\
P19 & {Research Engineer} & 3-4 years  & yes & yes  & yes \\
P20 & {Product Manager} & 5-9 years  & yes & yes  & yes \\
P21 & {Product Manager} & 5-9 years  & yes & yes  & no \\
P22 & {Researcher}  & 1-2 years  & yes & yes  & no \\
P23 & {Product Manager} & 10+ years  & yes & no & no \\
P24 & {Product Manager} & 10+ years  & yes & yes  & yes \\
P25 & {Research Manager}  & 5-9 years  & yes & yes  & yes  
\end{tabular}
\caption{Study participants and their responses to screener questions about prior experience working on AI systems, working on responsible AI projects, and working on accessibility projects, respectively.}
\label{table:participants}
\end{table*}

\subsection{Interview Protocol}
Each participant took part in a one-hour, semi-structured, remote interview. In general, the interviews followed critical incident technique \cite{flanaganCriticalIncidentTechnique1954} as we asked participants to walk through 1-3 examples of current and past projects. 
We asked participants to first summarize the project’s goals, stakeholders that they worked with, challenges that they encountered during the project, and its outcome. While not all projects directly involved accessibility or disability, we encouraged the participants to highlight issues related to those topics that came up at the time, as well as opportunities where accessibility and disability might have been relevant even if it was not explicitly discussed. After discussing projects, we asked participants to describe their prior training (formal or informal) they had received related to accessibility practices, and share any suggestions they had for addressing any challenges they encountered related to disability inclusion.

\subsection{Data Analysis}
Interviews were audio and video recorded. Transcripts of the discussions were initially generated using the video platform’s built-in AI transcription feature, which the research team then corrected and anonymized.

We used an inductive coding approach for qualitative thematic analysis \cite{clarkeThematicAnalysis2017,byrneWorkedExampleBraun2022}. In the initial round of analysis, each researcher read through at least two transcripts and wrote preliminary codes. The research team then met to discuss these codes. Following these discussions, two of the authors then read the 25 transcripts and engaged in analytic memoing, so that each transcript was analyzed by both individuals. Upon completing this across all transcripts, the two researchers met weekly to engage in reflexive thematic analysis sessions, while discussing interpretations of the data, and generating themes \cite{campbell2021reflexive}. They aggregated and shared themes with the rest of the study team for deliberation and rounds of iterative review. 
\section{Findings}
\subsection{Challenges at the Intersection of AI and Disability}
Across roles, AI practitioners discussed a wide variety of ways they encountered and interacted with disability in their work. While our investigation did not attempt to provide a comprehensive account of the issues surrounding AI and disability in the company’s products, we did identify several classes of problems that our participants recounted, including both traditional UI accessibility problems as well as issues specific to AI.
\subsubsection{Manifestations of Disability in AI}
Problematic representations of PWD in AI output have previously been documented by users, community members, and researchers  \cite{gadirajuWouldntSayOffensive2023,venkitAutomatedAbleismExploration2023,mackTheyOnlyCare2024}. However, given challenges our participants report in involving PWD in their work, our participants instead shared how they perceive the potential issues that AI may engender for PWD. Some participants reported noticing these types of issues in products that they had worked on. P7, a safety analyst, noted issues related to distortions in how PWD were represented by image models:

\begin{quote} for the [disabled] person it showed additional disability…  like his hand is like this but there is one more hand[s] [appearing] over here … this kind of issue came only for the [disabled] person.  \end{quote}

P7 went on to note that PWD were not represented by default, and only appeared when the prompt included disability, while other aspects of identity were included by default.

Other representational issues included stereotypical images and inspiration porn, where PWD are portrayed as exceptional to inspire others \cite{youngImNotYour1402326644}.  P25 described another stereotype perpetuated through text-to-image generation model they had tested:

\begin{quote} When you would prompt [the model] to say create an image ... [the model] would create a picture of an angry old man in a wheelchair and an offensive image. I mean all disabled people are not angry … we're not all old white men. we're not all in wheelchairs. Yeah obviously bad data training on their model. \end{quote}

P19, a research engineer with a machine learning (ML) background, explored how inspiration porn manifested in the company’s LLMs. He hypothesized that the general positivity and agreeableness of the models (an intentional design feature) led to inspiration porn in model output:

\begin{quote}I think the inspiration porn was a huge just overwhelming problem in [the model] because they have this just very strong baseline positivity—that kind of thing.  \end{quote}

In investigating these phenomena, P19 identified potential ways to mitigate these issues but was unsuccessful in implementing changes. In discussing the distribution of disability types in model output, P19 noted:

\begin{quote} The first thing I found when looking at this was basically a specific problem in a specific very early version of [the model] … where there was some specific training data that led the system to over-represent people with cerebral palsy and people in wheelchairs… so I was able to track that to where that was in the training data. Intervening on that is complicated. So I never tried to even intervene on that. \end{quote}
\subsubsection{Features That Don’t Work for Disabled Users}
In some cases, participants described working on projects that were not tested with PWD, or were not designed with them in mind. Because of this, they expressed doubt that the features would be accessible. 

Several participants raised concerns about the accessibility of projects that involved voice interfaces, for example, P2 noted “if [voice product] requires you to have an active conversation then it's very likely that someone with speech disabilities is probably not [able] to leverage this product.” Furthermore, they noted that these accessibility problems might not even be detected, as the accessibility problems may prevent giving feedback, noting “they [disabled users] will definitely not be able to surface feedback about this functionality.” We discuss this potential feedback gap further later in the findings.

Several participants also noted accessibility problems related to communication modality, such as interacting using sign language. P22 took a company-sponsored ASL class but encountered difficulty participating via video chat: “ there's a [video chat] feature … I'm impressed by the tech … it's the one that automatically shrinks [the video to frame your face]. I always had to turn that off for the ASL class because it just cuts your hand off.” Another participant, P11, noted that she was impressed by advances in speech recognition quality, but expressed disappointment about progress in sign language recognition: “the fact that we can't parse sign language. I'm like, how is this a thing?” Participants were regularly exposed to advancing capabilities of genAI which many were tasked with integrating into existing products. They were in turn, surprised when from their perspective, the extent of these capabilities was not being leveraged more widely to increase accessibility, such as through improving automatic sign language recognition.

While some of the above examples involve novel, emerging problems, in other cases the accessibility shortcomings were obvious, such as when standard accessibility features were overlooked due to aggressive launch schedules or gaps in staffing. P12 described one such case:

\begin{quote} Right now there's no captions for [voice product] and that is really an accessibility requirement. And then it's been like everyone went into a panic and [said] ‘how are we going to do this?’ … that caught people by surprise and it's like why does that catch people by surprise? because we don't have an accessibility representative who's able to oversee all these things and be like, "Hey everyone, remember we have to do this." \end{quote}

Deprioritized accessibility is a perennial problem requiring advocates’ persistence as new products ship. However, we juxtaposed it with the increasing pace at which genAI is being developed, exacerbating the potential for obvious regression of accessibility that accompany new releases.

\subsection{Lack of Knowledge and Access to Experts}
To address RAI issues, like those we shared earlier, companies and other organizations have developed processes for identifying, assessing, and mitigating potential risks or harms of AI \cite{microsoftResponsibleAIPrinciples}. However, such processes rely on organizational buy-in to implement them \cite{rakova2021responsible}. In this section, we discuss organizational factors that may impede AI teams’ ability to address potential AI issues for PWD.

All employees received some level of training about supporting users with disabilities (though the specifics of that training varies with the participant’s role and when they joined). All employees also had access to additional training on accessibility as needed. However, in practice, participants who were not in explicit accessibility roles generally lacked expertise in the company’s accessibility practices, particularly as they manifest in AI products. P11, a safety analyst, said “I don’t think I’m particularly knowledgeable.” P13, another safety analyst, said “I wish I would have been taught more about what's out there.” P25, a research manager, described his current project team thusly: “Our [project] team, the UX researchers, the UX engineers, all those people, I would say 80\% of them had no idea or experience at all with accessibility.” 

We should not expect that every team member will be an accessibility expert, as long as there are processes in place for flagging issues and routing them to experts. In this context, experts typically mean someone with skills to identify and mitigate inaccessibility would also have AI expertise. P14, an engineering manager leading a team on a new project using AI to increase  communication accessibility, described their ideal partner:

\begin{quote} So there's a lot of learning to speak the language and then how do we generate a meaningful data set? How balanced should that data set be? … There's a lot of those sorts of things we are having to learn and having to find resources has been really challenging and so we have to  …  go get a ML person who cares about accessibility to sit in the room with us. And then let's tell them that we actually want them to be opinionated. I don't want an ML person coming in and being like “you're the accessibility folks.” No, sure we'll talk about the accessibility pieces ourselves, but actually we need you to correct us on where we are doing things wrong. And that is awesome that we found enough people to be able to do that. but it maybe doesn't scale super great. \end{quote}

In practice, participants frequently described friction when finding or contacting experts when needed. A major cause for this friction was changes in organizational structure across the company. When asked who they would contact to discuss accessibility issues, P11, a safety analyst, said “I think a year ago I would have been able to answer this better and… right now things are just chaotic.” Several participants noted that contacts they had previously worked with had left their team or been reassigned—a challenge identified for RAI work more broadly \cite{ali2023walking}. P12, a UX manager, had identified an accessibility expert who could help with their team’s work, but then lost that contact due to organizational changes:

\begin{quote} In the beginning, it looked like they were going to be able to help because they were on [a related project] and we figured [it was] close enough. But then they got reorged and had layoffs. \end{quote}

P24, a product manager with more than 10 years of experience at the company, described how they had previously worked with centralized, company-wide accessibility experts, but noted that these resources seemed to have evaporated:

\begin{quote} I don't know that we have at the moment a direct contact specifically on accessibility … the team has changed over time. [The team] used to have a ‘contact us’ [link] or office hours or whatever, and we went through that process [previously], but I think the process changed and again I think it's been all self-service recently, of ‘here's some tools and templates’ but not necessarily an individual [contact]. \end{quote}

P18 noted that such changes in structure for review are simply part of how tech companies operate:

\begin{quote} All these tech companies will kind of pendulum swing between are we doing centralized responsibility [evaluation] or distributed responsibility … and so they moved a bunch of the responsibility work kind of distributed, [and] had propped up little mini versions of that in [product A] or [product B] so they kind of broke apart the team a little bit. \end{quote}
\subsection{Involvement of PWD and their Data}
The importance of including PWD in processes was shared across all participants. How PWD could be included varied greatly based on individual job roles. For data scientists and ML experts, these conversations generally revolved around access to data about, or from, PWD. For those in more user-centered roles, such as UX researchers, our discussions focused on including disabled users in testing to understand their firsthand experiences, as expressed by P16:

\begin{quote} It's kind of intuitive, if somebody who doesn't have sort of firsthand experience is building … some ML model to predict somebody's number of hours spent [using a product] … [they may fail] to take into consideration how some tasks might take some people much longer or the journey that somebody might go down. \end{quote}

Others, like P19, noted that including PWD was important for the sake of creating a spirit of participation:

\begin{quote} The things that are important to me from some of [this] work is the sort of participatory perspective, the value-based stance of saying people who are being represented in systems and by systems ought to have a say in shaping that. I think that's a sort of core thing that resonates with me. \end{quote}

\subsubsection{Data about Disability}
Some participants mentioned using datasets that included data labeled as coming from disabled people, or test cases related to disability, but noted that such datasets were rare. In some cases, participants thought (or suspected) that their datasets contained data about disability, but were unable to confirm this. P5, a data scientist, described a scenario in which they analyzed speech data but found that a subset of the data was unintelligible and could not be analyzed. They were told by a coworker that the data most likely included speech data from users with speech disabilities; with no confirmation of this fact, and no plan for how to handle such data, they simply excluded the unintelligible data from their analysis. P2, also a data scientist, postulated that some of the datasets they worked with included data from users with disabilities, but described the task of identifying this data as finding a “needle in a haystack.” 
\subsubsection{Working with Disabled Stakeholders}
Multiple participants expressed a desire to include more PWD in the process of developing and testing new technologies. They identified a number of barriers to doing so.

First, including participants with disabilities where they were not already involved incurred additional costs and required allocating time in the schedule to do so. P9, a UX manager, noted that disabled participants may be considered “niche” and thus required additional justification to be included in user testing:

\begin{quote} We have to justify the need to include certain niche participants… if we request them to be a part of the studies and it's for….wanting to be more inclusive, these populations are smaller and it's rightfully so we don't want to over sample or overburden those people. I think that most researchers don't [recruit PWD unless they are doing specific research that is specifically focused on that. \end{quote}

Second, finding participants with the needed characteristics could be difficult. Recruiting disabled populations is generally challenging \cite{lysaghtParticipantRecruitmentStudies2016,beckerRecruitingPeopleDisabilities2004} and this can be magnified for specific projects, especially when AI is involved. For example, P10 described their team’s efforts to include PWD in their work, but noted that their existing policy often meant that only a few disabled participants would be recruited:

\begin{quote} [Participant groups] are going to be really small … I think we aim for 10\% of those who identify with a disability and that is a broad spectrum … and it is a self ID, so … that means in a qualitative recruit where we want 20 participants we're only talking about two [people]. \end{quote}

Our participants identified an emerging challenge, which is that disability categories, and screening questions used to recruit participants, may not map to the specific populations needed for some AI-related projects. P11 described a previous project in which an image generation model created problematic outputs when representing people with limb differences. Because the company has generally settled upon a standard screener that does not include much detail (P10: "we've essentially curated [recruitment] down to three questions, which is not exhaustive at all"), P11 and their team considered conducting internal recruiting: “maybe we can find a group of folks who have prosthetic limbs, knowing that that's going to be really freaking hard, even in a company this big,” but were not able to identify such a group. 

Third, policies that were intended to support PWD could make it difficult to recruit disabled participants. P9, a UX researcher, described an experience in which they tried to include disabled stakeholders, but were ultimately unsuccessful:

\begin{quote} Several years ago, we wanted to do some research in the internal IT support space, specifically with users using assistive technology devices … we wanted to look into it and figure out what are the improvements that we can make to the products and services to better serve that population. We went through a pretty hefty process to try to get that research approved to launch with that population, and ultimately that process ended up being so lengthy that the research was never done because by the time the process to get approval [completed], the ship had sailed and the team just had bigger more important priorities by that time. So it did not happen. So that was really disappointing for me to see. \end{quote}

Participants navigated legal and review processes put in place to protect company workers and external participants in regard to what personal data is shared and how it is handled and stored. However, these protective policies were implemented such that the time required often exceeded that which product groups could pause to diversify their sources of user feedback.
\subsection{Conflicting Review Processes}
A common theme across our participants’ accounts was the challenge in navigating the various processes needed for approval and launch of any product. The company has had internal accessibility guidelines for many years; these generally include a series of guidelines (similar to WCAG) with required and optional criteria. It is generally expected that individual teams will manage evaluating products according to these guidelines, and that any launched projects will meet the minimum guidelines; exemptions from these requirements are usually temporary and require approval from senior leadership. Notably, these guidelines are generally written for the context of traditional graphical UIs and applications on mobile, desktop, and web platforms, rather than AI-based features or products, and are written at a very granular level, identifying specific problems, their impacts on users, and how to fix them.

More recently, the company has adopted a separate set of guidelines for issues pertaining to “responsible AI”—adherence to these guidelines is typically reviewed by a centralized team. As with accessibility guidelines, it is expected that any launched projects will meet the requirements, and exceptions to these guidelines require approval from senior leaders. In contrast to the accessibility guidelines, which are updated at a regular interval, these RAI guidelines (and their associated review processes) have changed more rapidly, and in fact were in the process of being updated as we conducted our interviews. In their current form, these guidelines are generally higher-level, and thus must be interpreted to be applied to any specific project or issue.

As noted in the previous sections, our participants identified a variety of problems that could negatively impact users with disabilities. Some of these issues involved traditional inaccessible UI elements, such as missing captions or alt text; these issues were generally well represented in the company’s accessibility guidelines. Other issues, such as AI model output that perpetuates negative stereotypes about PWD, may seem to fit more within the scope of RAI guidance. However, in practice we found that this distinction was often unclear to participants, and there was generally no process for resolving this ambiguity for any specific issue.

One problem that was raised by multiple participants involved with RAI review was that existing RAI processes and documentation often did not focus on PWD. During our interview with P5, they shared an internal document used in RAI review, but noted “the [safety] part, which is the one that maybe relates the most, doesn't have too many examples on accessibility and disability.” P7 noted that disability “was never a focus area … when we test, we focus on gender, we focus on race, we focus on religion, we focus on all but we never focused on disability.” P18, a researcher, described their process for working with teams to identify potential safety issues, and noted that disability was often not prioritized by their partner teams:

\begin{quote} When I have … been in roles to try to help lay out … the policies we care about [and note] we have things in there around representation or stereotypes and I'm trying to kind of lay out a plan, [I’ll say] why don't we look at this from a disability lens in addition to kind of the main things that we go for, gender, race; usually no attention is given or it's kind of crossed off just out of … resourcing might be an excuse that's given. \end{quote}

Across these discussions, we found that RAI evaluations typically focused on a specific set of identity categories that included race, gender, and sometimes age and nationality—while disability was never explicitly excluded, it was often noted that it was not a priority. P13, a safety analyst, also noted this pattern, while connecting it back to foundational work in RAI:

\begin{quote} We were working with a team … they were dealing with [a project] relative to biometrics, being able to detect a human within either an image or a video … and I think I made a comment about ‘we see here that you've done some testing relative to gender, relative to different skin tones and race, relative to age, to make sure there's similar levels of performance across different potential user groups.’ and then I pointed out, ‘have you done that relative to potential users with disabilities or physical bodily differences? And I think they were a bit confused at the question because I think sometimes people more think about race and gender before they think about disability ... It's out there, but I feel like it's not top of mind as much for people. \end{quote}

A second challenge shared by participants was that issues related to disability sometimes surfaced during RAI reviews, but that it was not always clear which issues should be “blockers,” or whose responsibility it was to flag or resolve these issues. As P11 noted:

\begin{quote} I would say that when an accessibility issue is particularly salient, we call it out. But as an example, for these [product] launches, we didn't say ‘make sure this works with a screen reader’, which is a gap. And also what does particularly salient mean? There's no clear line. And so it's like, does this look extra bad to me? Okay, let me flag it. \end{quote}

A third type of challenge involved disagreements between what would be prescribed by RAI policy and the expectations of providing equal access to all users. Generating descriptions of people for accessibility purposes can sometimes clash with other design priorities \cite{stanglPersonShoesTree2020,bennettItsComplicatedNegotiating2021}. P13, a safety analyst, shared a similar problem with respect to RAI policies:

\begin{quote} There's a team wanting to launch image descriptions, but it's where … we can use generative AI capabilities [so that] low vision and blind users could ask questions about the image …  And so I had some questions about their policy because they're using [a safety classifier] which blocks inputs and outputs that are considered inappropriate and that could include pornographic content, right? Or something like that. And so I guess my thing with that is that our policy is that…  it should block that content. But from an accessibility standpoint, that is not allowing content that's otherwise seen by able-bodied users to be viewed by …  users with disabilities …  And so my thing is like if we're allowing able-bodied users to see that content, then we should also allow disabled users to get a screen overview of that content …  I think we should be going against [responsible AI] policy in this scenario because it's not accessible. \end{quote}
\subsection{Negotiating Responsibility}
As noted previously, the ambiguity between what counts as “accessibility or disability exclusion” vs. “a RAI problem” introduces ambiguity that may cause some problems to fall through the cracks. At a more fundamental level, even when a problem is identified and understood, there may be disagreement about who is responsible for solving the problem—a common shortcoming of RAI practices, as noted by Widder and Nafus (\citeyear{widderDislocatedAccountabilitiesAI2023}). This tension was described by P25:

\begin{quote} All of the AI accessibility research projects that I've been working on the last few years always go back to the fundamental question. Who's paying for the UX study, the resources, the data, who has the headcount, where's the money going to come from? And my answer to them is why aren't you originally budgeting for this instead of treating it as something that you're not going to touch at first? You have an aging group of the population, the disability community. Are you not considering them part of your target group? I think that's ridiculous, and I get that we have limited assets. I get that people are downsizing. Things are being cut. … Headcount isn't cheap. I get that, but you have to have that in the original planning instead of thinking that somebody else is going to pick up the tab. \end{quote}

This question of who will pay to solve the problem can be complicated by the interaction between hardware and software product teams, which is exacerbated by the division within companies’ organizational structure for generative AI models and AI-based applications. P12, who served as UX manager for a software product, described her team’s discussions with an associated hardware team:

\begin{quote}[The hardware team] would come to us and say, "Who's going to fix this accessibility issue …?" And we're like, " it's not us." And we were getting a lot of push back for good reason. and so we were like, "How are we going to staff this? What are we going to do?" And that's I think when it started to get more pushed out that we need somebody to kind of handle this … It was like a hot potato. No one [on our team] was like, "No, that's on the platform”. No, that's on the hardware. That's who owns it. \end{quote}

This interconnection between multiple teams occurred in many examples, sometimes between three or more collaborating teams and their clients. For example, P1’s team developed a content creation application that leveraged a generative AI model (developed by a second team) which was then delivered to external customers:

\begin{quote} So now there are three different stakeholders involved. One is …. our team which has built out the model but … we don't build [the base model] … So, one of the questions I've seen is who actually owns [this content]? So, if something goes wrong … let's say that [the model] generates more people of color than others, then who owns [this problem], right? Who is responsible for it? \end{quote}

With the rise of genAI, product teams are being tasked with integrating foundation models made by other teams into their product features. The responsibility of resourcing accessibility compliance is long-contested—just two of the extreme possibilities are being included in a product’s resourcing needs and being resourced from a centralized accessibility budget. But incorporating genAI models that are developed by multiple far-flung teams without clear guidance on how inaccessibility is promulgated downstream compounded confusion about where responsibility lies. Lacking guidance, some participants were unable to address issues of disability exclusion, as the solutions were outside of their team’s control and were not anticipated when resourcing for their specific products.
\subsection{Solving Problems with Constrained Resources}
Many of our participants validated the importance of addressing AI’s impacts on PWD. As described previously, our participants often encountered barriers due to resource constraints, including limited time, money, headcount, or access to subject matter experts. Often, they perceived a gap in the work they were able to do and what they saw as needed. At the end of her interview, P11 commented, “actually the reason I signed up for this research is … I want to make it super clear that we need to do better here.”

Despite these limitations, participants shared examples of how they were able to solve problems amidst constrained resources. In particular, participants identified several important components to achieving success.

\subsubsection{Filling Gaps} 
In some cases, participants identified a gap in team structure or practice that was causing friction. One strategy in these situations was to simply fill the gap informally, and in doing so, demonstrate to leadership the need for a more permanent solution. This strategy was exemplified by P12, who identified a critical gap where a new AI product team was missing accessibility expertise; informally took on that role for several months until a role was formally created:

\begin{quote} So I was one of the first UXers on board. I was able to actually do things like revise our [requirements doc] and there was nothing about accessibility in there. So I added, “Hey we need to make sure that our first launch … passes [mandatory accessibility requirements] at least. And then there wasn't anyone really assigned to accessibility. So I just spent time trying to [fix that] ... I was able to get some resources from [another team] to do some test plans and testing for us. I got some engineers on board, started trying to get some designers  …  I was trying to just raise support and mention things that we should definitely cover for our first launch. 

So for six months … I built up more of my knowledge around how the accessibility world works at [our company] … Basically [I just kept] poking people and being like, “hey, hey, don't forget the [accessibility]”. And they'd be like, “what's that?” … [later] our lead program manager … was like, "Hey… we think we might be able to swing having a little bit more resourcing for accessibility." … I think basically people were just tired of me banging on about accessibility all the time. \end{quote}

\subsubsection{Volunteer Work} 
Several participants described additional work that they performed, outside of the scope of their official role, to make the company’s AI products more inclusive. Much of this work took place through an informal, company-wide program for “red teaming” models. This team tested models from the perspective of multiple identity categories, including disability, and recruited members from these diverse identity groups. Another impactful volunteer initiative is the company-wide affinity group for disabled employees. This group has advanced support for disability inclusion in the company’s products by collecting and organizing feedback from its many members, meeting with high-level stakeholders, and advocating for accessibility goals to be included in company-wide objectives and key results (OKRs).

\subsubsection{Creating Buy-In from Leadership} 
Participants noted that getting buy-in from company leaders was practically necessary to affect change within the company. P25, who worked at the company for over 10 years, stated that work “used to be very much more bottom up in terms of directives, but now it's kind of becoming a little bit more top down.” P25 noted that getting buy-in from executives was crucial to the success of their recent projects, and that key to this was framing accessibility projects as creating a “curb cut” effect:

\begin{quote} [Executives] were very interested in focusing on how we can use accessible AI and that whole concept to also have a curb cut effect eventually. So it's an innovative solution that looks like it's been designed for an accessibility purpose, but ends up having broad use among all users…so it's not just something that's used for disabled people. \end{quote}

\subsubsection{Balancing Accessibility with Speed} 
Ultimately, many of our participants saw value in advancing their work, including work at the intersection of AI and accessibility, but hoped to find the right balance between progress and risk. This balance was described eloquently by P14:

\begin{quote} The responsible AI piece is really the tricky part, and I think to some degree [our product] shipped and it's not the most accessible tool, we're still dealing…but I think we need to figure out ways to make that these things moving fast and be accessible, not in opposition, because as long as we think we see accessibility as sort of slowing us down, we're leaving people behind. and I think that that's sort of counter to [the company's] mission. \end{quote}
\section{Discussion}
\subsection{How AI is Impacting Accessibility}
Our study builds upon prior work in identifying an emerging set of concerns around how AI-powered systems may impact PWD. Our participants noted potential problems that appeared in the context of their work, including user interface accessibility issues, stereotypes and representation problems, and accessibility issues specific to emerging interaction modalities such as voice. We see value in continuing to catalogue these issues, and to explore their impacts both from the perspective of end users, as well as understanding how these problems manifest within organizations creating the technology.

In addition to identifying new categories of accessibility problems, our participants noted how the “AI gold rush” \cite{greensteinAIGoldRush2023} can impact PWD in other ways. Our participants described how their work was impacted by time pressure, reduced resources, staff reductions, and company reorganization. As AI is changing how software is created, we must continue to grow our understanding of how the operations of software organizations may impact users with disabilities. One notable example here is how the increasing complexity of software products, and the organizations that create them, result in multiple stakeholders who may have differing opinions about where problems lie and whose responsibility it is to address them: for example, an accessibility bug that appears on a mobile phone which is running an application that is accessing a generative model may be mitigated in multiple ways.

Finally, participants noted that access to information about users with disabilities remains a perennial problem. This includes the traditional difficulties in recruiting and collecting feedback from disabled participants, a growing need for collecting feedback from specific user groups (e.g., people with particular health conditions or physiological characteristics), and the challenge of creating datasets that represent the diversity of PWD.
\subsection{Category Confusion}
A persistent theme across many of our interviews was a general lack of clarity about the boundaries between categories such as fairness, inclusion, RAI, and accessibility. 

Specifically, we identify the importance of understanding \textbf{AI harms that impact people with disabilities}—including problems that disproportionately affect disabled people or problems that impact those groups differently. Notably, multiple participants stated resources and processes related to responsible AI frequently focused on a specific set of identity categories (race and gender, sometimes age and ethnicity) that often overlooked disability. We see an urgent need to incorporate our understanding about disability inclusion into mainstream RAI resources and curricula. Furthermore, large companies should perhaps examine their processes for triaging problems at the intersection of AI and disability, to ensure that these problems are not overlooked.

While we see clear value in teaching AI practitioners to understand the needs of users with disabilities, we also see an opportunity to develop career paths at the union of UX, accessibility, and RAI. Across different roles, our participants raised concerns that could be partially addressed through education—there is an opportunity to teach RAI and AI safety practitioners more about disability, and to teach UX and accessibility professionals more about AI. There may also be opportunities to develop career paths at the intersection of these fields; much as accessibility practitioners sometimes act as specialized versions of UX practitioners, we imagine opportunities to create professionals who are themselves experts at disability inclusion in AI systems.
\subsection{Practitioners’ Knowledge Gaps about Disability}
Throughout our interviews, participants suggested various types of information resources that might help them address the needs of PWD in their work. While our interviews did not go into sufficient depth to answer the specifics of what information resources should be created, we found a consistent set of knowledge gaps that may be addressed. 

First, participants said that they would benefit from having guidelines, checklists, and other procedural support to ensure that their work would be verifiably inclusive of PWD. Checklists and similar resources are often discussed in the context of responsible AI practices \cite{lecaResponsibleAISoftware2024, madaioAssessingFairnessAI2022}; in this context, these resources could be extended to cover challenges experienced by PWD when using AI. Second, participants noted that they had little knowledge of the experiences of PWD, which made it difficult to anticipate how PWD might use or be impacted by a system; while this challenge may be addressed in part by increasing the diversity of project teams \cite{baldassarreFosteringHumanRights2025}, there may be opportunities to educate practitioners to better understand PWD, and thereby more easily predict negative impacts. Finally, as noted previously, participants did not always have a clear understanding of how to resolve problems that they did identify, which may be addressed both through information resources and policy changes.
\section{Limitations and Future Work}
One limitation of this work is that our participants worked at a single organization. Thus the experiences of our participants were shaped to some degree by the culture of the overall company, which may include company-wide policies or communications. However, we did recruit participants across a variety of job roles, product areas, and with various amounts of time spent at the company. In some cases, participants mentioned other work that they had performed before joining the company—while we did not include these examples in our analysis, we note that some participants had developed their perspective from experiences at multiple organizations. This limitation suggests an opportunity for future work to expand this investigation across companies with different organizational structures, sizes, and cultures.

A second limitation of the work is that we were typically only able to collect data from one member on a particular team; thus we were generally unable to triangulate and report experiences across multiple perspectives. Focusing future studies on a particular team could provide a richer context to the issues encountered, though at the possible cost of presenting less breadth of experiences.

Finally, we note that in casting a wide net in the job roles that we considered as participants, we did not recruit members from a specific job role with similar goals, backgrounds, or job requirements. This was intentional, in that we expected (and found) that challenges of disability inclusion appeared across various job roles. However, this makes it more difficult to generalize between participants. Future research may wish to examine the impacts of AI on specific job roles such as UX researchers or accessibility analysts.
\section{Conclusion}
Growth in AI, and the pervasiveness of AI among software systems, raises new questions about how to ensure that products are usable by, and accessible to PWD. Our study shows that these problems can appear in unexpected ways beyond the traditional scope of user interface accessibility. This research suggests the need for reevaluating education and training around RAI and accessibility, revising launch priorities to ensure that PWD are considered as valuable stakeholders throughout the process, and continuing to expand our knowledge base on how AI impacts PWD.


\bibliography{aaai25}
\clearpage

\end{document}